\documentclass[conference]{IEEEtran}
\usepackage{cite}
\usepackage{amsmath,amssymb,amsfonts}
\usepackage{algorithmic}
\usepackage{graphicx}
\usepackage{textcomp}
\usepackage{xcolor}
\def\BibTeX{{\rm B\kern-.05em{\sc i\kern-.025em b}\kern-.08em
    T\kern-.1667em\lower.7ex\hbox{E}\kern-.125emX}}
\usepackage{graphicx}

\usepackage{pgfplots}
\usepackage{pgfplotstable}
\usepackage{booktabs, colortbl, array}
\usepackage{enumitem}
\usepackage{rotating}
\usepackage{balance}
\usepackage{bbm}
\usepackage{amssymb}
\usepackage{amsmath}
\usepackage{amsthm}

\usepackage{url}
\newtheorem{definition}{Definition}

\begin{document}
\title{Towards a Model for LSH}
\author{\IEEEauthorblockN{Li Wang}
\IEEEauthorblockA{\textit{Wufong University, China}
}}
\maketitle              % typeset the header of the contribution
\begin{abstract}
As data volumes continue to grow, clustering and outlier detection algorithms are becoming increasingly time-consuming. Classical index structures for neighbor search are no longer sustainable due to the "curse of dimensionality". Instead, approximated index structures offer a good opportunity to significantly accelerate the neighbor search for clustering and outlier detection and to have the lowest possible error rate in the results of the algorithms. Locality-sensitive hashing is one of those. We indicate directions to model the properties of LSH.
\end{abstract}

\section{Basic Definition}
\begin{figure}[t]
  \centering
  \includegraphics[width=0.5\columnwidth]{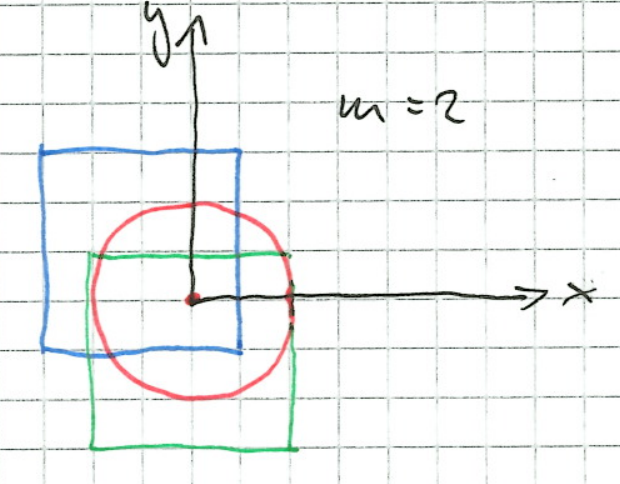}
  \caption{Basic Definition.}
  \label{fig:sigmoid}
\end{figure}
The idea is to consider the convex, Voronoi-like cells of k-means as a $d$-dimensional grid of hyper-cubic cells which have a volume such that an expectation of $n/k$ points are located within each. We denote the side length of these cells with $b$. The query is modeled as a ball with perimeter $s$, centered at origin. In the beginning, we assume that the ball is actually from a maximum metric such that the query is as well a hypercube with side length $s$.

In a locality sensitive hashing we use a number $m$ of alternative clusterings, and we refine for each of the $m$ clusterings exactly one cell, in which the query point (origin) is located. Therefore, the selectivity of the query response corresponds to the intersection of the union of all $m$ cells with the query ball. We start with the analysis of the case $s=b=1$. We assume that each of the $m$ cells is selected uniformly and independently such that the origin (query point) is located inside, so the upper boundary of each cell in each dimension is uniformly taken from the interval $[0..\tfrac{1}{2}]$ and the lower boundary is consequently from $[-\tfrac{1}{2}..0]$.

In the following  we will use several definite integrals which we denote in a bit unusual way which helps for clarity we will write
\[\int ... \mbox{ }(0\le x \le 1) \mbox{ instead of }\int_0^1 ... \mbox{ d}x.\]
We will need a few integrals throughout this paper:
\[\int x+\tfrac{1}{2}\mbox{ }(0\le x \le\tfrac{1}{2})  =  \tfrac{3}{8}\]
\[\int\hspace{-3mm}\int (x+\tfrac{1}{2})(y+\tfrac{1}{2})\mbox{ }(0\le x,y \le\tfrac{1}{2})  =  \tfrac{9}{64}\]
\[\int...\int (x_1+\tfrac{1}{2})\cdot ...\cdot (x_d+\tfrac{1}{2})\mbox{ }(0\le x_1, ..., x_d \le\tfrac{1}{2})  =  \Big(\tfrac{3}{8}\Big)^d\]
\[\int\hspace{-3mm}\int\min(x_1,x_2)+\tfrac{1}{2}\mbox{ }(0\le x_1, x_2 \le\tfrac{1}{2})  =  \tfrac{5}{24}\]
%\int\min(x_1,x_2,x_3)+\tfrac{1}{2}\mbox{ }(0\le x_1 \le\tfrac{1}{2})\mbox{ }(0\le x_2 \le\tfrac{1}{2})\mbox{ }(0\le x_3 \le\tfrac{1}{2}) & = & \tfrac{11}{128}\\
\begin{eqnarray*}\int...\int\min(x_1,...,x_d)+\tfrac{1}{2}\mbox{ }(0\le x_1, ... , x_d \le\tfrac{1}{2}) & = \\
=d\cdot \int x^{d-1}\cdot\big(x+\tfrac{1}{2}\big) \mbox{ }(0\le x \le\tfrac{1}{2})& = & \frac{2d+1}{(d+1)\cdot 2^{d+1}}
\end{eqnarray*}
\[\int\hspace{-3mm}\int y-v \mbox{ }(0\le y \le\tfrac{1}{2}, -\tfrac{1}{2}\le v \le 0) = \tfrac{1}{8}\]
\[\int ... \int (y_1-v_1)\cdot ... \cdot (y_d-v_d) \mbox{ }(0\le y_1, ..., y_d \le\tfrac{1}{2}, -\tfrac{1}{2}\le v_1, ..., v_d \le 0) = \Big(\tfrac{1}{8}\Big)^d\]
And, finally, we can also solve the following combination:
\[\int...\int \big(\min(x_1, ..., x_d)+\tfrac{1}{2}\big)\cdot(y_1-v_1)\cdot ... \cdot (y_m-v_m) \mbox{ }(0\le x_1, ..., x_d, y_1, ..., y_m \le\tfrac{1}{2},-\tfrac{1}{2}\le v_1, ..., v_m \le 0)\]
which gives:
\[=\frac{2m+1}{8^d\cdot(m+1)\cdot 2^{m+1}}.\]
\begin{definition}
$p(m,\ell,d)$ is the (hyper-) volume of the hypercube representing the query which is occupied by at least $\ell$ of the $m$ cells (with $\ell\le m$). Here the $m$ cells are uniformly selected in $\mathbb R^d$ such that the query point is inside the cell.
\end{definition}
Note that $p(m,1,d)$ corresponds to the selectivity of the hashing provided that $b$ and $s$ are of equal size.

For the case $m=\ell=1$ we can easily derive the closed formula of $p(1,1,d)$. We simply have to form, in each quadrant, the (identical) expectation with which the query is occupied by the cell:
\[p(1,1,d) = 2^d\cdot \int...\int (x_1+\tfrac{1}{2})\cdot ... \cdot (x_d+\tfrac{1}{2})\mbox{ }(0\le x_1, ..., x_d \le\tfrac{1}{2})  =  \Big(\tfrac{3}{4}\Big)^d\]
The case $m>1, \ell=1$ can be reduced to the case $m=1$:
\[p(m,1,d) = \bigg({m\atop 1}\bigg)\cdot p(1,1,d) - \bigg({m\atop 2}\bigg)\cdot p(1,2,d) + \bigg({m\atop 3}\bigg) \cdot p(1, 3, d) - ... \pm \bigg({m\atop m}\bigg)\cdot p(1, m-1, d)\]

For the case $m=1, \ell=2$, we start with the analysis of $d=1$ and construct this case with two variables $x$ and $y$. We want to estimate the area which is covered by $x$ and $y$, which is
\[p(1,2,1) = \int\hspace{-3mm}\int\left\{\begin{array}{ll}1-|x-y|&\mbox{ if }x<0\mbox{ xor } y<0\\
1-\max(|x|,|y|)&\mbox{ otherwise}\end{array}\right.(-\tfrac{1}{2}\le x,y \le\tfrac{1}{2})=\frac{7}{12}.\]
This is an independent information in each dimension such that
\[p(1,2,d) = p(1,2,1)^d = \Big(\frac{7}{12}\Big)^d.\]
\begin{figure}[t]
  \centering
  \includegraphics[width=0.8\columnwidth]{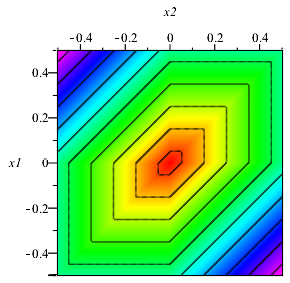}
  \caption{Two-dimensional Integral.}
  \label{fig:bla}
\end{figure}
For a generalization to $\ell=3$ we make a similar case distinction:
\[p(1,3,1)=\int\hspace{-3mm}\int\hspace{-3mm}\int \left\{\begin{array}{ll}
    1-\max(x,y,z)&\mbox{if }x,y,z\ge 0\\
    %1-|z-\max(|x|,|y|)|&\mbox{if }x,y<0\mbox{ and } z\ge 0\\
    1-\max(x,y)+z&\mbox{if }x,y\ge 0\mbox{ and } z< 0\\
    ...
\end{array}\right.(...) = \frac{15}{32}.\]
where the first case occurs in 2 octants and the second case in the remaining 6 octants. We solve the first case by letting
\[2\cdot \int\hspace{-3mm}\int\hspace{-3mm}\int 1-\max(x,y,z) = 6\cdot \int (1-x)\cdot x^2 = \tfrac{5}{32}\]
and similarly the second case
\[6\cdot \int\hspace{-3mm}\int\hspace{-3mm}\int 1-\max(x,y)+z = 12\cdot \int\hspace{-3mm}\int (1-x +z)\cdot x =\tfrac{5}{16}.\]

For the general case, we consider $\ell$ variables $x_1, ..., x_\ell$. Again we consider a solution space where we distinguish if each variable is greater or less than 0. All variables are equivalent, thus we consider all quadrants of the solutions space equally where the same number of variables is $<0$. There is a number of\[\left({\ell \atop i}\right)\mbox{ quadrants of the solution space having $i$ out of $\ell$ variables} <0.\]
Therefore, we have
\[p(1,\ell,1) = \sum_{0\le i\le \ell}\left({\ell \atop i}\right)\cdot \int...\int 1-\max(x_1, ..., x_i)+\min(x_{i+1}, ..., x_\ell)\hspace{5mm}(0\le x_1, ..., x_i\le \tfrac{1}{2}, -\tfrac{1}{2}\le x_{i+1}, ..., x_\ell\le 0)\]

\section{Related Work}

There are various approaches for \textbf{different data types}. The data can be of any type, as long as a distance function exists. Fixed-length text data often uses Hamming distance \cite{DBLP:journals/tjs/HoOK18} and the similarity between variable length text is often measured using the edit distance  \cite{DBLP:journals/pvldb/XiaoWL08}. A common measure for set data is the Jaccard distance  \cite{DBLP:conf/sigmod/DengT018,DBLP:journals/tods/XiaoWLYW11}, whereas the similarity of documents is processed with cosine-like similarity measures \cite{DBLP:series/synthesis/2013Augsten,DBLP:conf/icde/ShangLLF17}.

\textbf{Approximate} nearest neighbor search techniques can also be applied to the similarity join problem, however without guarantees on completeness and exactness of the result. There may be false positives as well as false negatives. Recently an approach \cite{DBLP:journals/tkde/YuNLWY17} to Locality Sensitive Hashing (LSH) is used on a representative point sample, to reduce the number of lookup operations. LSH is of interest in theoretical foundational work, where a recursive and cache-oblivous LSH approach \cite{DBLP:journals/algorithmica/PaghPSS17} was proposed. The topic of approximate solutions for the similarity join is also an emerging field in deep learning \cite{DBLP:journals/corr/abs-1803-04765}. There are approximative approaches which target low dimensional cases (spatial joins in 2--3 dimensions \cite{DBLP:conf/icde/BryanEF08}) or higher (10--20) dimensional cases  \cite{DBLP:conf/focs/AndoniI06}. Very high-dimensional cases, with dimensions of $128$ and above have been targeted with Symbolic Aggregate approXimation (SAX) techniques \cite{DBLP:journals/concurrency/MaJZ17}) to generate approximate candidates. SAX techniques rely on several indirect parameters like PAA size or the iSAX alphabet size.

There are preconstructed indexing techniques, which are based on \textbf{space-filling curves} and applied to the similarity join problem. Specifically, where the data is sorted efficiently with respect to one or more Z-order curves \cite{DBLP:conf/kdd/DittrichS01, DBLP:journals/tkde/KoudasS00, DBLP:conf/icde/LiebermanSS08} in order to test the intersection of the hypercubes in the datastructures. Others propose space-filling curves, to reduce the storage cost for the index \cite{DBLP:journals/tkde/ChenGLJC17}.
%GESS \cite{DBLP:conf/kdd/DittrichS01} and
LESS \cite{DBLP:conf/icde/LiebermanSS08} targets GPUs and not multi-core environments. ZC and MSJ \cite{DBLP:journals/tkde/KoudasS00} as well as the SPB-tree index \cite{DBLP:journals/tkde/ChenGLJC17}, although simple, they require space transformations and preprocessing, which make them hard to parallelize.

\textbf{EGO family} of $\epsilon$-join algorithms. The EGO-join algorithm is the first algorithm in this family introduced by B\"{o}hm et al. in  \cite{epsilongridorder}. The Epsilon Grid Order (EGO) was introduced as a strict order (i.e. an order which is irreflexive, asymmetric and transitive). It was shown that all join partners of some point $\mathbf x$ lie within an $\epsilon$-interval, of the Epsilon Grid Order. Algorithms of the EGO family exploit this knowledge for the join operation. The EGO-join has been re-implemented as a recursive variant with additional heuristics, to quickly decide whether two sequences are non-join-able \cite{DBLP:conf/dasfaa/KalashnikovP03}. Further improvements proposed two new members of this family, the EGO$^{*}$ \cite{DBLP:journals/is/KalashnikovP07} algorithm and its extended version called Super-EGO \cite{DBLP:journals/vldb/Kalashnikov13} target multi-core environments using a multi-process/multi-thread programming model. Super-EGO proposes a dimensional reordering \cite{DBLP:journals/vldb/Kalashnikov13}. In the experiments Super-EGO encounters some difficulties with uniformly distributed data, particularly when the number of data objects exceeds millions of points or the dimensionality is above $32$.%\vspace{1.0mm}

% Besides the EGO family, the Quickjoin \cite{DBLP:journals/tods/JacoxS08} is another non-index based algorithm, which solve the similarity join problem. There has been also a parallel implementation proposed, which is called the MRSimJoin \cite{DBLP:conf/sigmod/SilvaR12, DBLP:conf/cloudi/SilvaRT12}. Code \footnote{http://www.public.asu.edu/~ynsilva/SimCloud/SJSurvey/}

If the similarity join runs multiple times on the same instances of the data, one might consider \textbf{index-based approaches} \cite{DBLP:conf/icde/BohmK01, DBLP:journals/jda/ParedesR09, DBLP:journals/tkde/ChenGLJC17}, such as R-tree \cite{DBLP:conf/sigmod/BrinkhoffKS93} or \textit{M}-tree \cite{DBLP:conf/vldb/CiacciaPZ97}. Index-based approaches have the potential to reduce the execution time, since the index stores pre-computed information that significantly reduces query execution time. This pre-computational step could be costly, especially in the case of List of Twin Clusters (LTC)
\cite{DBLP:journals/jda/ParedesR09}, where the algorithm needs to build joint or combined indices for every pair of points in the dataset. The D-Index \cite{DBLP:journals/mta/DohnalGSZ03} and its extensions (i.e. eD-Index \cite{DBLP:conf/dexa/DohnalGZ03} or i-Sim index \cite{DBLP:conf/sisap/PearsonS14}) build a hierachical structure of index levels, where each level is organized into separable buckets and an exclusion set. The most important drawback of D-Index, eD-Index and i-Sim is that they may require rebuilding the index structure for different $\epsilon$.%However, non of the mentioned spatial indexing approaches is parallelized.
% However, spatial indexing approaches do not perform well in high-dimensional spaces, due to the ``curse of dimensionality'' \cite{DBLP:conf/stoc/IndykM98}.

\textbf{Data partitioning across multiple machines} is not the main focus of this paper where we assume that the data fits into main memory. The case of relational join algorithms has been studied extensively in the past \cite{DBLP:conf/sigmod/SchneiderD89, DBLP:conf/kdd/WangMP13, DBLP:journals/pvldb/FierABLF18}.
The similarity join has been successfully applied in the distributed environment with different MapReduce variants \cite{DBLP:conf/waim/LiWU16, DBLP:conf/sigmod/McCauley018, DBLP:journals/pvldb/FierABLF18}. Another distributed version is proposed in \cite{DBLP:conf/sigmod/ZhaoRDW16}. There, a multi-node solution with load-balancing is used, that does not require re-partitioning on the input data. This variant focuses on minimization of data transfer, network congestion and load-balancing across multiple nodes.

The similarity join has been already implemented for \textbf{Graphics Processing Units (GPUs)}.
In \cite{DBLP:conf/btw/BohmNPZ09} the authors use a directory structure to generate candidate points. On datasets with 8 million points, the proposed GPU algorithm is faster than its CPU variant, when the
 $\epsilon$-region has at least 1 or 2 average neighbors.
LSS \cite{DBLP:conf/icde/LiebermanSS08} is another similarity join variant for the GPU, which is suited for high dimensional data. Unfortunetly both \cite{DBLP:conf/icde/LiebermanSS08} and \cite{DBLP:conf/btw/BohmNPZ09}  are targeted to NVIDIA GPUs and have been optimized for an older version of CUDA.

\subsection{Cache-oblivious Algorithms}
\noindent Cache-oblivious algorithms \cite{DBLP:conf/focs/FrigoLPR99} have attracted considerable attention as they are portable to almost all environments and architectures. Algorithms and data structures for basic tasks like sorting, searching, or query processing \cite{DBLP:conf/sigmod/HeLLY07} and for specialized tasks like ray reordering \cite{DBLP:journals/tog/MoonBKCKBNY10} or homology search in bioinformatics \cite{DBLP:journals/bmcbi/FerreiraRR14} have been proposed. Two important algorithmic concepts of cache-oblivious algorithms are localized memory access and divide-and-conquer. The Hilbert curve integrates both ideas. The Hilbert curve defines a 1D ordering of the points of an 2-dimensional space such that each point is visited once. Bader et al. proposed to use the Peano curve for matrix multiplication and LU-decomposition \cite{DBLP:conf/para/BaderM06, DBLP:conf/europar/Bader08}. The algorithms process input matrices in a block-wise and recursive fashion where the Peano curve guides the processing order and thus the memory access pattern. In \cite{loopsjournal}, cache-oblivious loops have been applied to K-means clustering and matrix multiplication. %We considerably improve memory locality and runtime by introducing the Fast General Form (FGF-) Hilbert curve (see Section \ref{sec:hilbert}).
% FGF-Hilbert is an improvement to our FUR-Hilbert curve \cite{DBLP:conf/bigdataconf/BohmPP16, IEEE:transbigdata/BohmPP18}, which has been applied to matrix multiplication, K-means, Cholesky decomposition and the algorithm by Warshall.

\subsection{Optimized Techniques for Specific Tasks or Hardware}
\noindent The library BLAS (Basic Linear Algebra Subprograms) \cite{DBLP:journals/toms/DongarraCHD90} provides basic linear algebra operations together with programming interfaces to C and Fortran. BLAS is highly hardware optimized: specific implementations for various infrastructures are available, e.g. ACML for AMD Opteron processors or CUBLAS for NVIDIA GPUs. The Math Kernel Library (MKL) contains highly vectorized math processing routines for Intel processors. These implementations are very hardware-specific and mostly vendor-optimized. Moreover, they are designed to efficiently support specific linear algebra operations. Experiments demonstrate that the cache-oblivious approach reaches a performance better than BLAS on the task of the similarity join for points of dimensions in the range of $\{2,...,64\}$.
\section{Conclusion}
This survey reviews several approaches of HPDM from many research groups world wide. Modern computer hardware supports the development of high-performance applications for data analysis on many different levels. The focus is on modern multi-core processors built into today's commodity computers, which are typically found at university institutes both as small server and workstation computers. So they are deliberately not high-performance computers. Modern multi-core processors consist of several (2 to over 100) computer cores, which work independently of each other according to the principle of ``multiple instruction multiple data'' (MIMD). They have a common main memory (shared memory). Each of these computer cores has several (2-16) arithmetic-logic units, which can simultaneously carry out the same arithmetic operation on several data in a vector-like manner (single instruction multiple data, SIMD). HPDM algorithms must use both types of parallelism (SIMD and MIMD), with access to the main memory (centralized component) being the main barrier to increased efficiency.

\nocite{DBLP:conf/sigmod/BohmFP08,DBLP:conf/icdt/BerchtoldBKK01,DBLP:conf/icdm/BohmK02,10.1007/BFb0000120,DBLP:conf/icde/BohmOPY07,
DBLP:conf/cikm/BohmBBK00,DBLP:journals/jiis/BohmBKM00,DBLP:conf/adl/BohmBKS00,DBLP:conf/edbt/BohmK00,
DBLP:journals/sadm/AchtertBDKZ08,DBLP:journals/bioinformatics/BaumgartnerBBMWOLR04,DBLP:journals/jbi/BaumgartnerBB05,
DBLP:conf/icdt/BerchtoldBKK01,DBLP:conf/edbt/BohmP08,DBLP:conf/kdd/AchtertBKKZ06,DBLP:conf/cikm/BohmFOPW09,
DBLP:conf/ssdbm/BohmPS06,DBLP:conf/kdd/BohmHMP09,DBLP:conf/dawak/BerchtoldBKKX00,DBLP:conf/sdm/AchtertBDKZ08,
DBLP:journals/jdi/BaumgartnerGBF05,DBLP:journals/kais/MaiHFPB15,DBLP:journals/tlsdkcs/BohmNPWZ09,
DBLP:conf/dexa/BohmK03,DBLP:conf/icde/BohmGKPS07,DBLP:conf/miccai/DyrbaEWKPOMPBFFHKHKT12,DBLP:conf/kdd/PlantB11,
DBLP:journals/bioinformatics/PlantBTB06,DBLP:conf/icdm/ShaoPYB11,DBLP:conf/icdm/PlantWZ09,
DBLP:journals/tkdd/BohmFPP07,DBLP:conf/pakdd/BohmGOPPW10,DBLP:conf/btw/BohmNPZ09,DBLP:conf/icdm/MaiGP12,
DBLP:journals/kais/ShaoWYPB17,DBLP:conf/kdd/YeGPB16,DBLP:conf/kdd/FengHKBP12,DBLP:journals/envsoft/YangSSBP12,
DBLP:conf/icdm/GoeblHPB14,DBLP:conf/kdd/AltinigneliPB13,DBLP:conf/icdm/YeMHP16,DBLP:conf/kdd/Plant12,
DBLP:conf/kdd/Plant12,DBLP:conf/cikm/BohmBBK00,DBLP:journals/kais/BohmK04,DBLP:conf/icdm/BohmK02,
DBLP:conf/icdt/BerchtoldBKK01,loopsjournal,Bially1969SpacefillingCT,
Prusinkiewicz:1986:GAL:16564.16608,10.1007/BFb0000120,DBLP:conf/icdm/BohmK02}
\bibliographystyle{unsrt}
\bibliography{bibliography}
\end{document}